\documentclass[aps,prl,twocolumn,superscriptaddress,showpacs,
floatfix,preprintnumbers]{revtex4-1}

\usepackage{soul}
\usepackage{color}
\usepackage{amssymb}
\usepackage{graphicx}
\usepackage[normalem]{ulem}

\begin{document}
\title{Connecting Neutron Star Observations to Three-Body Forces \\
in Neutron Matter and to the Nuclear Symmetry Energy}

\author{A.~W. Steiner}
\affiliation{Joint Institute for Nuclear Astrophysics, National
Superconducting Cyclotron Laboratory and the \\ Department of Physics and
Astronomy, Michigan State University, East
Lansing, MI 48824}
\affiliation{Institute for Nuclear Theory, University of 
Washington, Seattle, WA 98195}

\author{S. Gandolfi}
\affiliation{Theoretical Division, Los Alamos National Laboratory, Los
Alamos, NM 87545}

\begin{abstract}
Using a phenomenological form of the equation of state of neutron
matter near the saturation density which has been previously
demonstrated to be a good characterization of quantum Monte Carlo
simulations, we show that currently available neutron star mass and
radius measurements provide a significant constraint on the equation
of state of neutron matter. At higher densities we model the equation
of state using polytropes and a quark matter model, and we show that
our results do not change strongly upon variation of the lower
boundary density where these polytropes begin. Neutron star
observations offer an important constraint on a coefficient which is
directly connected to the strength of the three-body force in neutron
matter, and thus some theoretical models of the three-body may be
ruled out by currently available astrophysical data. In addition, we
obtain an estimate of the symmetry energy of nuclear matter and its
slope that can be directly compared to the experiment and other
theoretical calculations.
\end{abstract}


\maketitle

\emph{Introduction:} While experimental information on matter near the
nuclear saturation density is plentiful, there are only a few
experimental constraints on matter above the saturation density and,
when available, they are contaminated by strong systematic
uncertainties. The variation in the energy of nuclear matter with
isospin asymmetry is particularly uncertain, since laboratory nuclei
probe only nearly isospin-symmetric matter. There is a strong effort
in trying to constrain the symmetry energy from intermediate energy
heavy-ion collisions~\cite{Tsang:2009}, giant resonances in
nuclei~\cite{Carbone:2010}, and parity-violating electron-nucleus
scattering~\cite{Horowitz:2001,Roca-Maza:2011}.

Theoretical computations of neutron-rich matter are also difficult,
owing to the poor quality of effective forces in dealing with
neutron-rich matter~\cite{Gandolfi:2011b} and uncertainties in the
nature of the three-neutron force. At low densities, neutron matter is
well understood because the two-body neutron--neutron interaction is
constrained by experimental scattering phase shift data. At higher
densities, three different classes of methods have emerged for
computing the properties of neutron matter. The first class is based
on phenomenological forces like the Skyrme
interaction~\cite{Stone:2007}. The second class of calculations is
based on microscopic nuclear Hamiltonians that typically include two-
and three-body forces obtained from chiral effective field theories,
adjusted using renormalization group techniques to do perturbative
calculations~\cite{Hebeler:2010}. However the renormalization of the
nuclear Hamiltonian induces many-body forces that have been carefully
included in light nuclei~\cite{Jurgenson:2011} but not yet in nuclear
matter. An alternative approach is the Brueckner-Hartree-Fock
theory, which has been extensively used to study nuclear matter
and hyperonic matter~\cite{Li:2008,Vidana:2011,Schulze:2011}. The
third class uses nuclear potentials, like Argonne and Urbana/Illinois
forces, which reproduce two-body scattering and properties of
light nuclei with very high precision~\cite{Wiringa:1995,Pieper:2001}.
In the latter case, the interaction is designed to have small
non-local terms, giving the potentials a hard core. The calculations
can be performed in nonperturbative framework, and the strong
correlations are solved by using correlated wave functions. The ground
state of nuclear systems is determined by using the
cluster-expansion~\cite{Akmal:1998} or using quantum Monte Carlo (QMC)
methods. QMC methods have proven to be a very powerful tool to accurately
study properties of light nuclei~\cite{Pudliner:1997,Pieper:2008} and
nuclear matter~\cite{Gandolfi:2010}. All three of these classes suffer
from strong uncertainties above the saturation density, both regarding
the method and the nuclear Hamiltonian.

On the other hand, astrophysical observations of neutron star masses
and radii probe the equation of state (EOS) of dense, neutron-rich matter
above the saturation density. Two types of neutron star mass and
radius measurements have provided for progress on constraining the
EOS: the measurement of the general relativity-corrected radiation
radius of quiescent low-mass x-ray binaries (qLMXBs)~\cite{Rutledge:1999}, and
the observation of photospheric radius expansion bursts which provides
a simultaneous measurement of both the mass and
radius~\cite{vanParadijs:1979,Steiner:2010,Ozel:2010}.
Reference~\cite{Steiner:2010} has demonstrated that these two sets of data
provide significant constraints on the EOS, ruling out several
currently available theoretical models of dense matter.

In this work, we show that these astrophysical observations are
beginning to constrain the nature of the three-body force in neutron
matter. We construct a phenomenological description of the EOS near
the saturation density which faithfully reproduces QMC simulations of
neutron matter and can represent a wide range of EOSs at high
density~\cite{Gandolfi:2009,Gandolfi:2011}. Utilizing the currently
available astrophysical observations, we show that two parameters,
closely connected with the strength of the three-body force and
related to the magnitude and density dependence of the symmetry
energy, are constrained by the observational data.

\emph{The model:} For densities below about half the saturation
density, neutron stars consist of a crust which is solid except for a
thin shell at the surface. Since the uncertainty in the EOS of the
crust leads to an error in the radius which is much smaller than the
current observational uncertainty, we ignore variations in the EOS of
the crust~(see, e.g., \cite{Lattimer:2001}). We use the outer crust from
Ref.~\cite{Baym:1971} and the inner crust from
Ref.~\cite{Negele:1973}. Near and above the saturation density, we use
a parametrization of neutron matter with the form
\begin{equation}
\label{eq:eosnm}
\epsilon_{\mathrm{NM}}=\rho\left[a\left(\frac{\rho}{\rho_0}\right)^\alpha
+b\left(\frac{\rho}{\rho_0}\right)^\beta+m_n\right] \,,
\end{equation}
where $\rho$ is the nucleon number density, $m_n$ is the nucleon mass,
and $a$, $\alpha$, $b$ and $\beta$ are free parameters. In
Ref.~\cite{Gandolfi:2009,Gandolfi:2010} it has been shown that this
general form accurately fits the EOS of pure neutron matter given by
QMC calculations using realistic nuclear Hamiltonians including two-- and
three--body forces. The uncertainty of the fit is much smaller than
that from the three-body force. In the neutron matter case, the two
parameters $a$ and $\alpha$ are mostly related to the nucleon--nucleon
force, while the parameter $b$ is mostly sensitive to the
corresponding symmetry energy $E_{\mathrm{sym}}$. The parameter
$\beta$ is sensitive to the particular model of three-neutron force
(see Ref.~\cite{Gandolfi:2011}). The range of parameters which
subsumes all reasonable QMC calculations of neutron matter is 
$12.7<a<13.3$ MeV, $0.48<\alpha<0.52$, $1<b<5$ MeV and $2.1<\beta<2.5$.

Because the parametrization in Eq.~\ref{eq:eosnm} describes neutron
matter, we must also make a small correction to the neutron matter EOS
due to the presence of a small number of protons. In order to estimate
this correction, we examine several Skyrme models from
Ref.~\cite{Stone:2003}, all chosen to have reasonable saturation
properties and symmetry energies sufficiently strong as to prevent
pure neutron matter from appearing in the maximum mass neutron star.
We compute the mean and root-mean-square deviation of the ratio of the
pressure of neutron matter to the pressure of neutron star matter as a
function of energy density over all the Skyrme models in our set. In
our fiducial model, we apply a randomly distributed correction to the
pressure with the same mean and root-mean-square deviation as that
obtained in the Skyrme models, to our neutron matter EOS. This
correction is 0.86$\pm$0.03 at saturation density. For comparison, we
also repeat our analysis without including the uncertainty in the
ratio of the pressures, simply applying the mean correction estimated
from the Skyrme models. We have also checked this correction is
similar to that given by similar relativistic mean-field models.

Above the saturation density, four-body forces, hyperons, Bose
condensates, and quark degrees of freedom may contribute to the EOS
and our parametrization will no longer be appropriate. We take
$\rho_{t}=0.40~\mathrm{fm}^{-3}$ as a reasonable upper limit for our
parametrization of neutron-rich matter. Our fiducial model describes
matter at higher densities by using a sequence of two
piecewise-continuous polytropes, $P=\varepsilon^{1+1/n}$ with
polytropic index $n$. The three parameters which describe the
high-density EOS are $n_1$ and $n_2$, the indices of the two
polytropes and $\varepsilon_P$, the transition energy density between
the two polytropes. Similar parametrizations have been used to
describe the EOS at high densities and can mimic the presence of phase
transitions.

Alternatively, we describe matter at high densities with a
parametrization of quark matter, with a polytrope at moderate densities
to represent the possible presence of a mixed phase. For the quarks we
use the model proposed by Alford et al.~\cite{Alford:2005}:
\begin{equation}
P=\frac{3a_4}{4\pi^2}\mu^4-\frac{3a_2}{4\pi^2}\mu^2-B \,,
\end{equation}
where $P$ is the pressure, $\mu$ is the quark chemical potential, the
coefficient $0.6<a_4<1$ describes corrections to the massless free
Fermi gas contribution from strong interactions, the coefficient
$a_2=m_s^2 - 4 \Delta^2$ subsumes corrections from quark masses
and color superconductivity, and $B$ is the bag constant. A largest
possible range for $a_2$ is between $(150~\mathrm{MeV})^2-4
(200~\mathrm{MeV})^2$ which corresponds to a bare strange quark with a
large quark gap and $(400~\mathrm{MeV})^2$, which corresponds to a
zero gap and strange quarks which receive significant contributions
from chiral symmetry breaking. This gives four parameters for the
high-density part: the index of the polytrope, $a_2$, $a_4$, and the
transition energy density between the polytrope and quark matter which
fixes the bag constant $B$.

To match our parametrization to the constraints from neutron star mass
and radius measurements, we use the method outlined in
Ref.~\cite{Steiner:2010}. To that original data set we add a recent
measurement from the transiently accreting neutron star
U24~\cite{Guillot:2011}. We also add the constraint that models must
be able to support at least a 1.93 solar mass neutron star, consistent
with the $1-\sigma$ lower limit from Ref.~\cite{Demorest:2010}. We use
Bayesian analysis, taking a uniform prior distribution for EOS
parameters and using marginal estimation to compute the posterior
probability distribution for EOS parameters, the EOS, and the mass
versus radius curve.

\begin{figure}
\parbox{1.65in}{
\vspace*{-3.1in}
\includegraphics[width=1.65in]{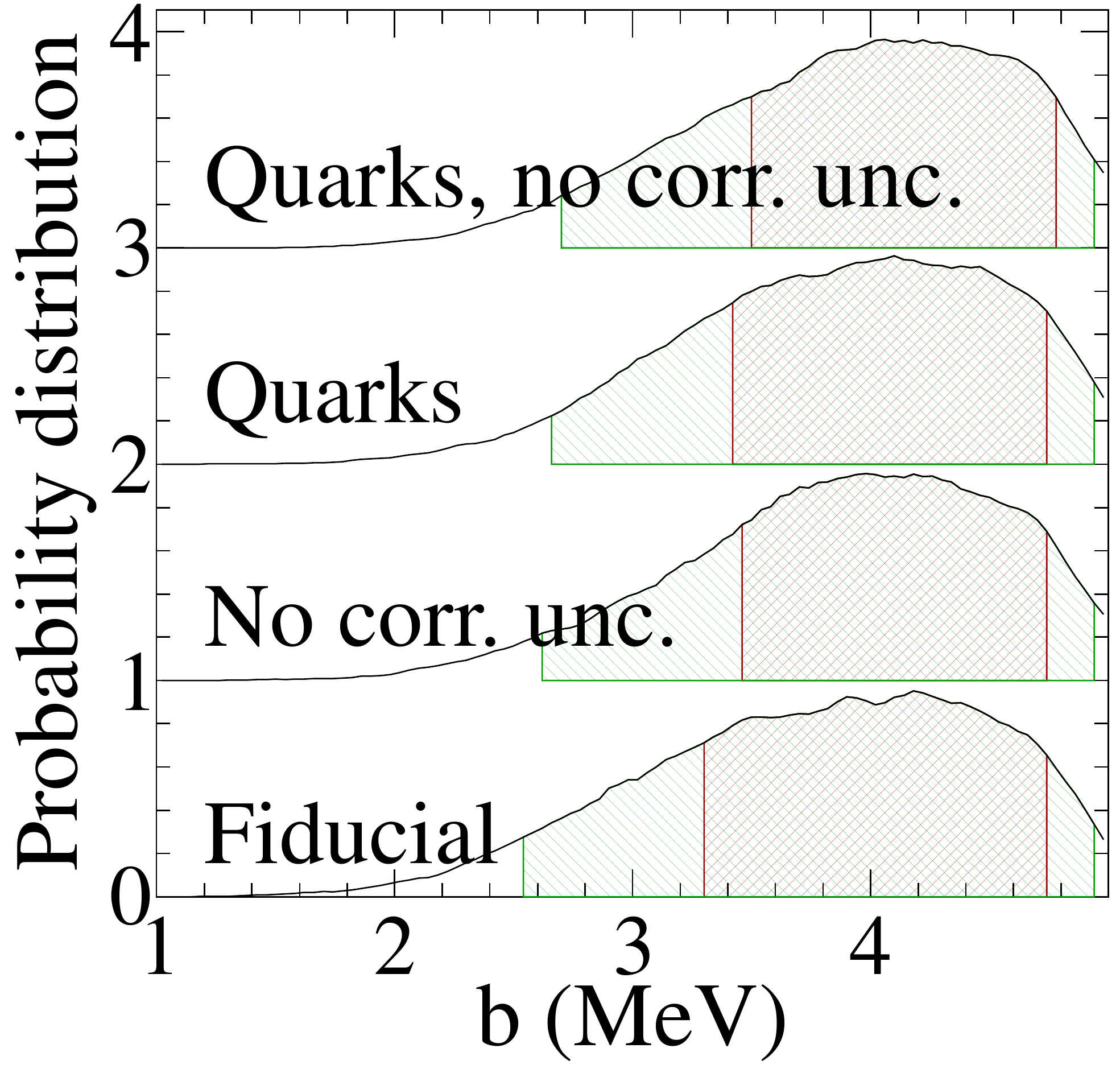}
\includegraphics[width=1.65in]{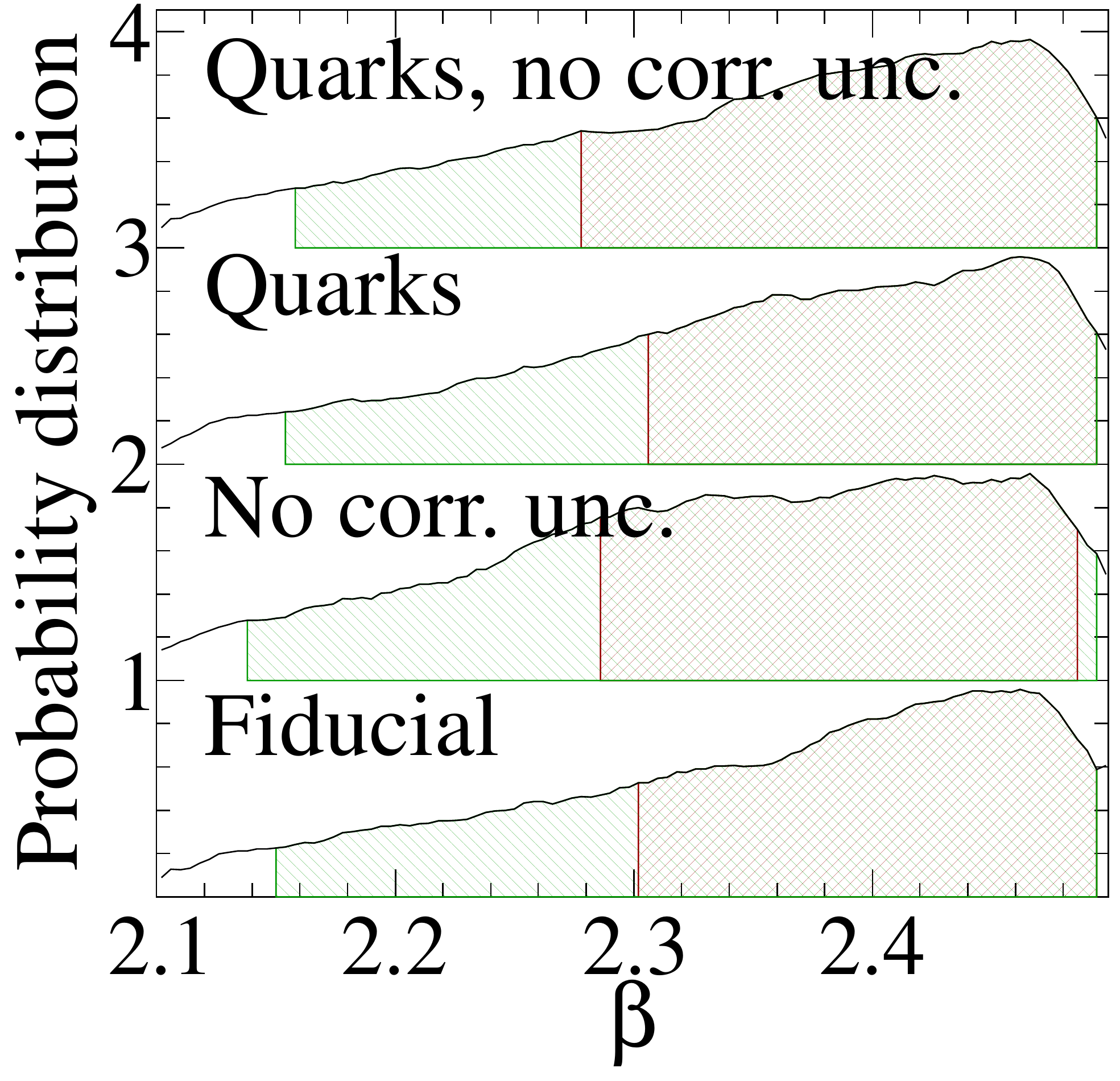}
}
\includegraphics[width=1.65in]{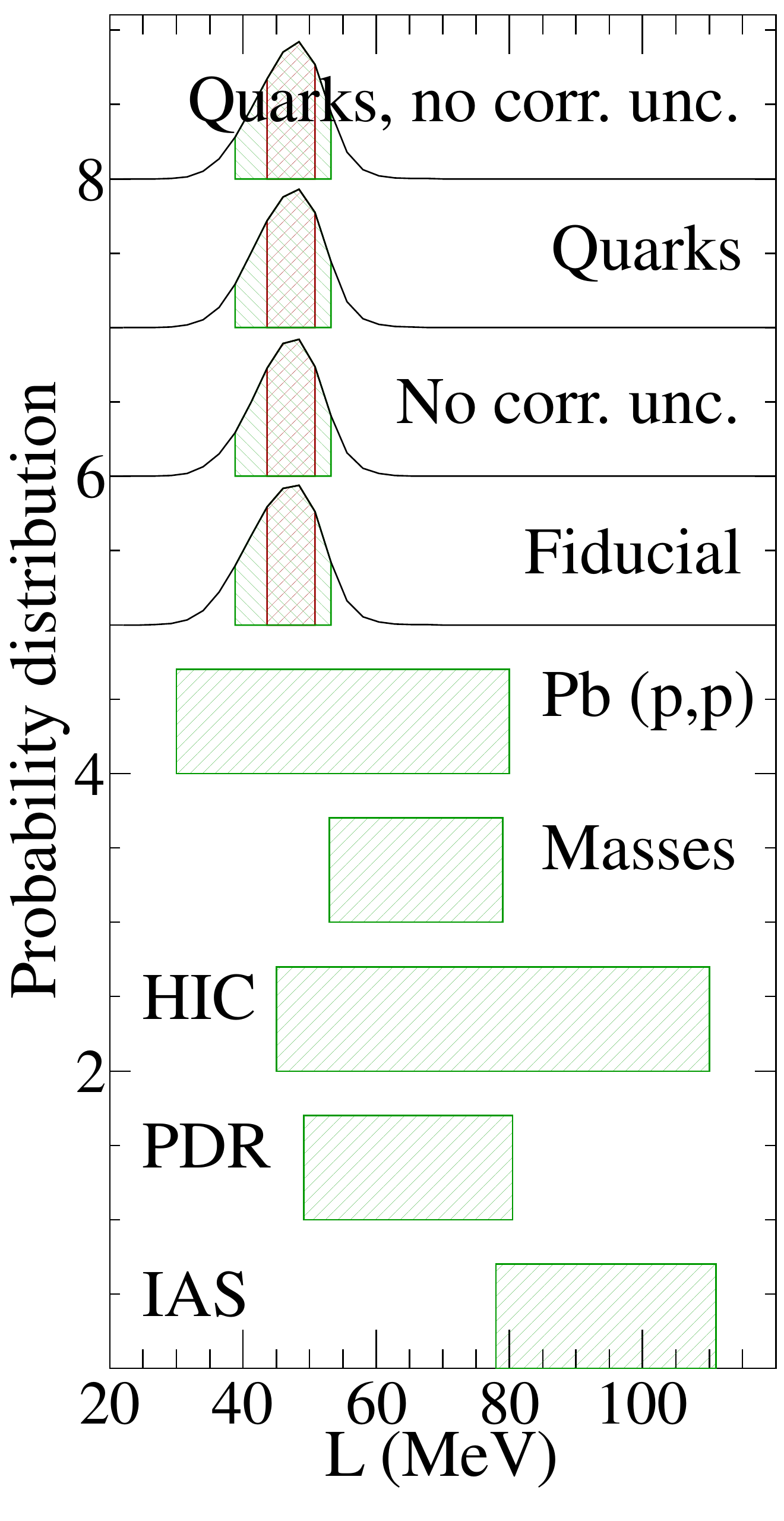}
\caption{(Color online) The probability distributions and 68\% (dark
  red areas) and 95\% (green areas) confidence ranges for the
  parameters $b$ and $\beta$, and the density derivative of the
  symmetry energy, $L$. All distributions have been rescaled so that
  their peak is unity and then vertically shifted by an arbitrary
  amount. We compare our predicted value
  of $L$ with constraints from nuclear masses (``Masses'')~\cite{Liu:2010},
  heavy ion collisions (``HIC'')~\cite{Tsang:2009},
  pygmy dipole resonances (``PDR'')~\cite{Carbone:2010}, isobaric
  analog states in nuclei (``IAS'')~\cite{Danielewicz:2009}, and
  antiprotonic atoms (``Pb(p,p)'')~\cite{Warda:2009}. The
  parametrization of the high-density part of the EOS is presented as
  Supplemental material~\cite{supplemental}. }
\label{fig:parms}
\end{figure}

\emph{Results and discussion:} Our constraints on the parameters $b$
and $\beta$ for the neutron matter EOS near the saturation density are
given in Fig.~\ref{fig:parms}, and the parametrization of the
high-density part of the EOS is presented as Supplemental
material~\cite{supplemental}. We find that the posterior probability
distributions for the parameters $a$ and $\alpha$ associated with the
two-body force are almost flat as expected because they are related to
the low-density part of the EOS. However, the parameters $b$ and
$\beta$ are strongly constrained by observations to ranges $3.3 < b <
4.8$ MeV and $2.28<\beta<2.5$, independent of the nature of the
high-density EOS. The results labeled ``no corr. unc.'' contain no
correction for the uncertainty in the ratio of the pressures of
neutron matter to neutron star matter in each panel and are nearly
indistinguishable from the results where this uncertainty is included
(and as a result we always include this uncertainty in the results
shown below). We have also checked that the effect of the varying the
matching density the neutron matter EOS of Eq.~\ref{eq:eosnm} with the
polytrope between $\rho_t=0.32$ fm$^{-3}$ and $0.48$ fm$^{-3}$ is
small.

Taking the nuclear matter binding energy at saturation to be $-$16
MeV, the nuclear symmetry energy is
$E_{\mathrm{sym}}=16~\mathrm{MeV}+a+b$. The density derivative of the
symmetry energy, $L\equiv 3 \rho_0 (dE_{\mathrm{sym}}/d\rho)_{\rho_0}$
is
\begin{equation}
L = 3 \left(a~\alpha +b~\beta \right)
\end{equation}
Taking the parameter ranges for $a$ and $b$ showed in Fig.
\ref{fig:parms} we find $32<E_{\mathrm{sym}}<34$ MeV and $43<L<52$ MeV
to within 68\% confidence. These tight constraints are possible
because of an interplay between the strong correlation in QMC
calculations of neutron matter between $E_{\mathrm{sym}}$ and $L$ and
the constraints on the EOS from neutron star radii. The correlation
between $E_{\mathrm{sym}}$ and $L$ results from a separation between
short- and long-distance parts of the three-neutron
force~\cite{Gandolfi:2011}. Neutron star mass and radius measurements
imply the radius is nearly independent of mass and relatively small,
which tends to select a smaller value for $L$ and the value for
$E_{\mathrm{sym}}$ is then constrained from the correlation. We also
find that the two- and three-body forces contribute almost equally to
the symmetry energy at the saturation density, but this demarcation is
more model dependent. Finally, our constraints on the neutron matter
EOS are consistent with and complimentary to those from heavy-ion
collisions~\cite{Danielewicz:2002}, which principally probe symmetric
matter.

\begin{figure}
\includegraphics[width=0.95\columnwidth]{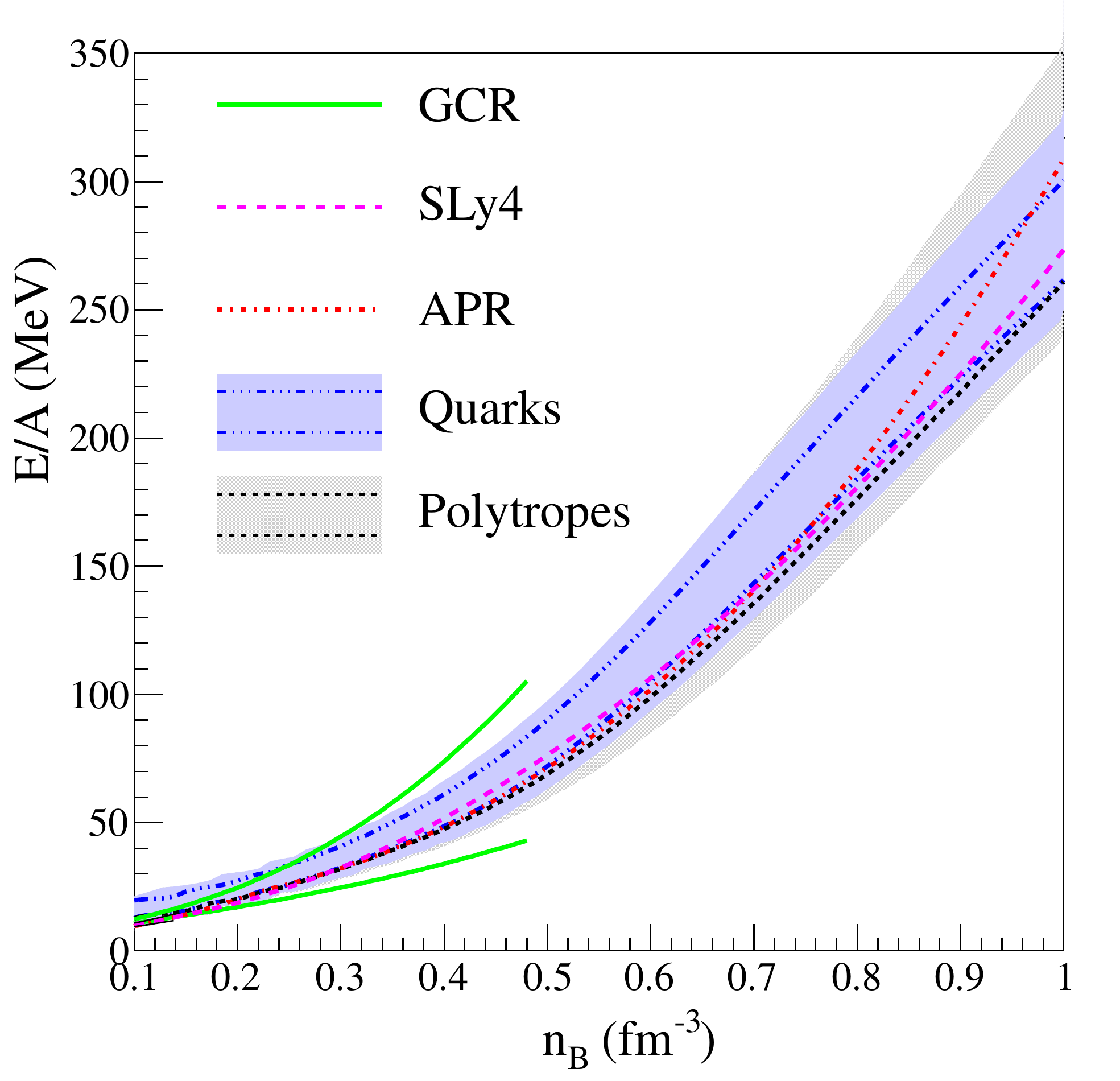}
\caption{(Color online) The 68\% (dashed and dot-dashed lines) and
95\% (shadowed areas) confidence ranges for the energy per baryon as a
function of the baryon density as constrained by the astrophysical
observations. The APR~\cite{Akmal:1998} and SLy4~\cite{Chabanat:1995}
EOSs are also plotted, as well as the limits obtained in
Ref.~\cite{Gandolfi:2011} (GCR).}
\label{fig:enb}
\end{figure}

The various models of three-body forces in neutron matter, that are
typically constrained in light nuclei~\cite{Pieper:2001}, and the fact
that the ranges of $b$ and $\beta$ are nearly independent of the high
density EOS implies that neutron star mass and radius measurements can
also constrain three-neutron forces. The ranges for $b$ and $\beta$
are smaller than the constraints determined from the wide range of
possible three-body forces given in Ref.~\cite{Gandolfi:2011},
demonstrating that the astrophysical observations are ruling out more
extreme models for the three-neutron force. The corresponding
EOSs are given in Fig.~\ref{fig:enb} along with the EOS
of Akmal, Pandharipande, and Ravenhall (APR)~\cite{Akmal:1998} and Skyrme model 
SLy4~\cite{Chabanat:1995}. The solid
lines show the limits obtained by Gandolfi, Carlson, and Reddy (GCR) in
Ref.~\cite{Gandolfi:2011}, obtained without any constraints from
neutron star observations.

\begin{figure}
\includegraphics[width=0.95\columnwidth]{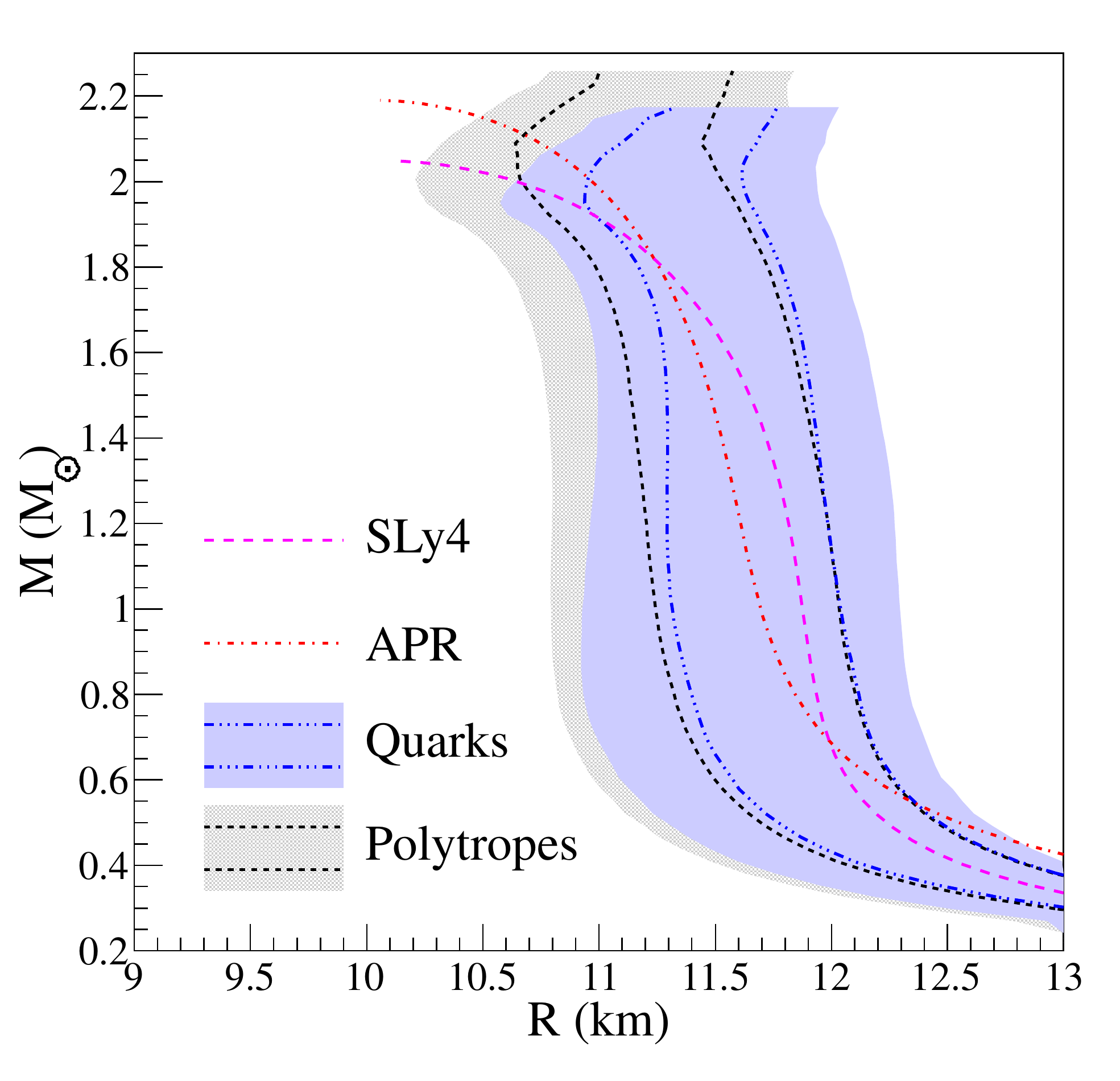}
\caption{(Color online) The mass-radius curves for the models considered in
this work. The range of radii for 1.4 solar mass neutron stars,
between 11 and 12 km, is similar to that obtained in
Ref.~\cite{Steiner:2010}. The mass-radius curves for APR~\cite{Akmal:1998} and
SLy4~\cite{Chabanat:1995} are also given. The labeling is the same as
in Fig.~\ref{fig:enb}.}
\label{fig:mvsr}
\end{figure}

In Fig.~\ref{fig:mvsr}, we show the probability of the mass of neutron
stars as a function of the radius for the different models. The quark
models have slightly larger radii and slightly smaller maximum masses,
but the general trend is similar to that described in
Ref.~\cite{Steiner:2010}. Neutron star radii lie between about 11 and
12.3 km regardless of mass to within 68\% confidence. As well as being
consistent with QMC calculations from Ref.~\cite{Gandolfi:2011} these
results are also consistent with the recent analysis of the EOS of
neutron matter using chiral effective theories in
Ref.~\cite{Hebeler:2010b}.

\begin{figure}
\includegraphics[width=0.95\columnwidth]{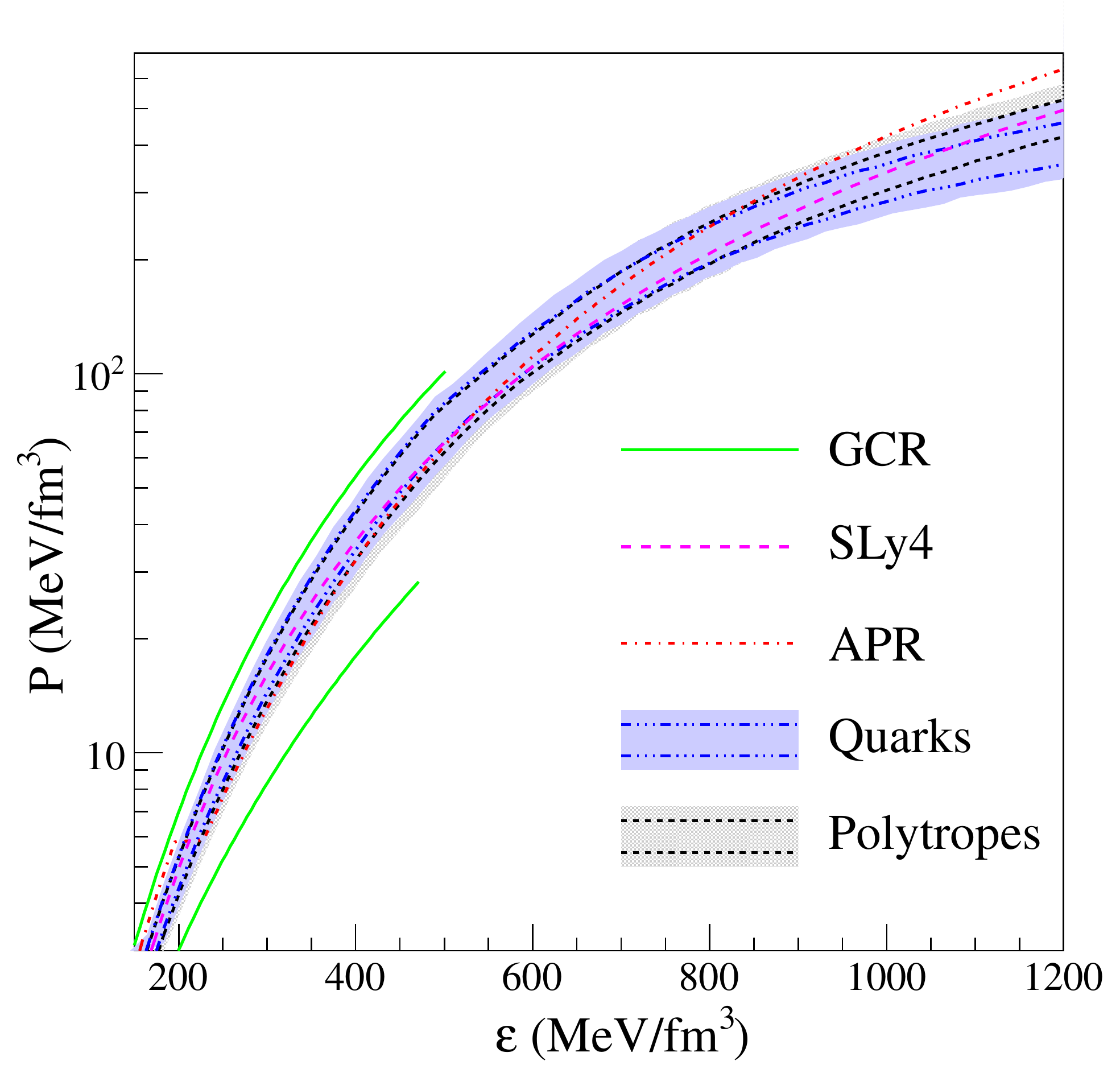}
\caption{The pressure as a function of the energy density.
The labeling is the same as in Fig.~\ref{fig:enb}.}
\label{fig:eos}
\end{figure}

Finally, we show the pressure of neutron star matter as a function of
the energy density in Fig.~\ref{fig:eos}. The results from
APR~\cite{Akmal:1998}, Sly4~\cite{Chabanat:1995} and
GCR~\cite{Gandolfi:2011}. The quark matter EOS is slightly softer at
higher energy densities, but these energy densities are often beyond
the central density of the maximum mass neutron star. The pressure in
APR is a bit larger than the observations suggest~\cite{Steiner:2010},
as is clear also in Fig.~\ref{fig:enb}. We stress that any EOS outside
the results presented in Figs.~\ref{fig:enb} and ~\ref{fig:eos}, like
APR at large energy densities, do not support constraints given by
neutron star observations (see also Ref.~\cite{Steiner:2010}).

\emph{Conclusions:} We find that neutron star mass and radius
measurements can be used to test and calibrate the three- and
many-body nuclear forces in the contest of dense infinite matter. In
particular we show that, when two- and three-body forces are
parametrized as a sum of power-laws in the baryon density, the
astrophysical observations strongly constrain both the coefficient and
exponent which describes the three-body force. We also find novel
constraints on the symmetry energy, driven partially by the strong
correlation between $S$ and $L$ obtained in QMC calculations of
neutron matter.

There are potential corrections which are not yet well understood,
including four-body forces, relativistic corrections, and the possible
presence of hyperons~\cite{Li:2008,Vidana:2011,Schulze:2011}. Our
model partially takes these into account through the high-density
polytropes (used above $\rho_t$) which are not constrained. However,
if these corrections are strong below $n_B=0.32~\mathrm{fm}^{-3}$,
then our constraints will have to be revisited accordingly.
Some hyperonic models are consistent with our
results~\cite{Bednarek:2011,Weissenborn:2011} while others do not
support neutron stars above 1.93 solar
masses~\cite{Vidana:2011,Schulze:2011}.

Another important difficulty is that there are several potential
systematic uncertainties in the neutron star mass and radius
measurements which are not yet under control. In the case of the photospheric radius expansion x-ray bursts, 
these include the nature of the relationship between the
Eddington flux and the point at which the photosphere returns to the
neutron star surface, the evolution of the spectrum during the tail of
the burst~\cite{Suleimanov:2011}, a modification of the spectrum due
to accretion, and violations of spherical symmetry. In the qLMXBs, the
X-ray spectra might contain high-energy power-law features not present
in the atmosphere models.

\emph{Acknowledgments:} We thank J. Carlson and S. Reddy for helpful
comments. A.W.S. is supported by Chandra Grant No. TM1-12003X,
by the Joint Institute for Nuclear Astrophysics at MSU under NSF PHY
Grant No. 08-22648, by NASA ATFP Grant No. NNX08AG76G, and
by DOE Grant No. DE-FG02-00ER41132. S.G. is supported by DOE
Grants No. DE-FC02-07ER41457 (UNEDF SciDAC) and No. DE-AC52-06NA25396.
This work was also supported
by the DOE Topical Collaboration ``Neutrinos and
Nucleosynthesis.''


%

\newpage

\section*{Supplemental Material}

\begin{figure*}[h]
\includegraphics[width=1.65in]{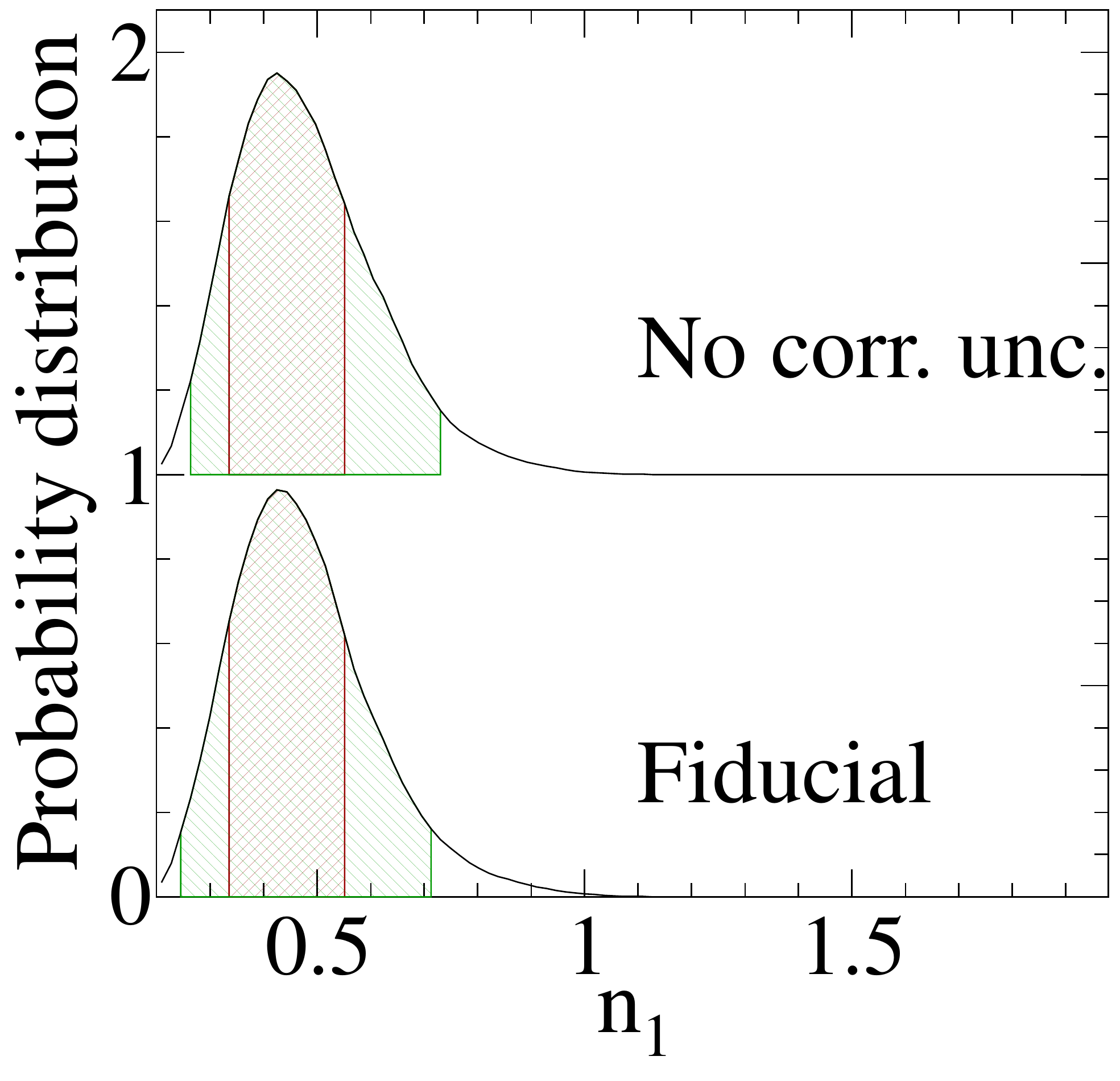}
\includegraphics[width=1.65in]{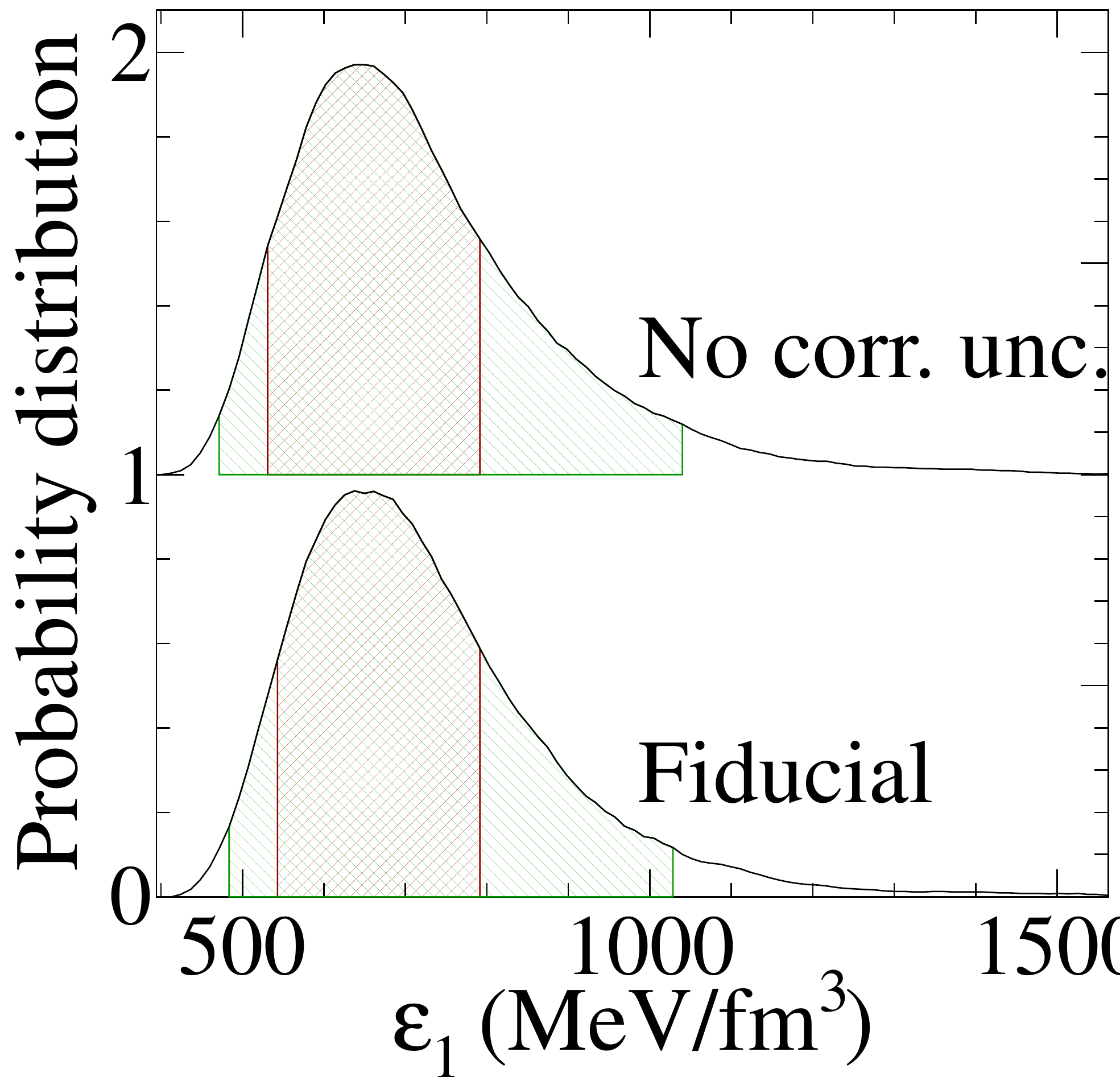}
\includegraphics[width=1.65in]{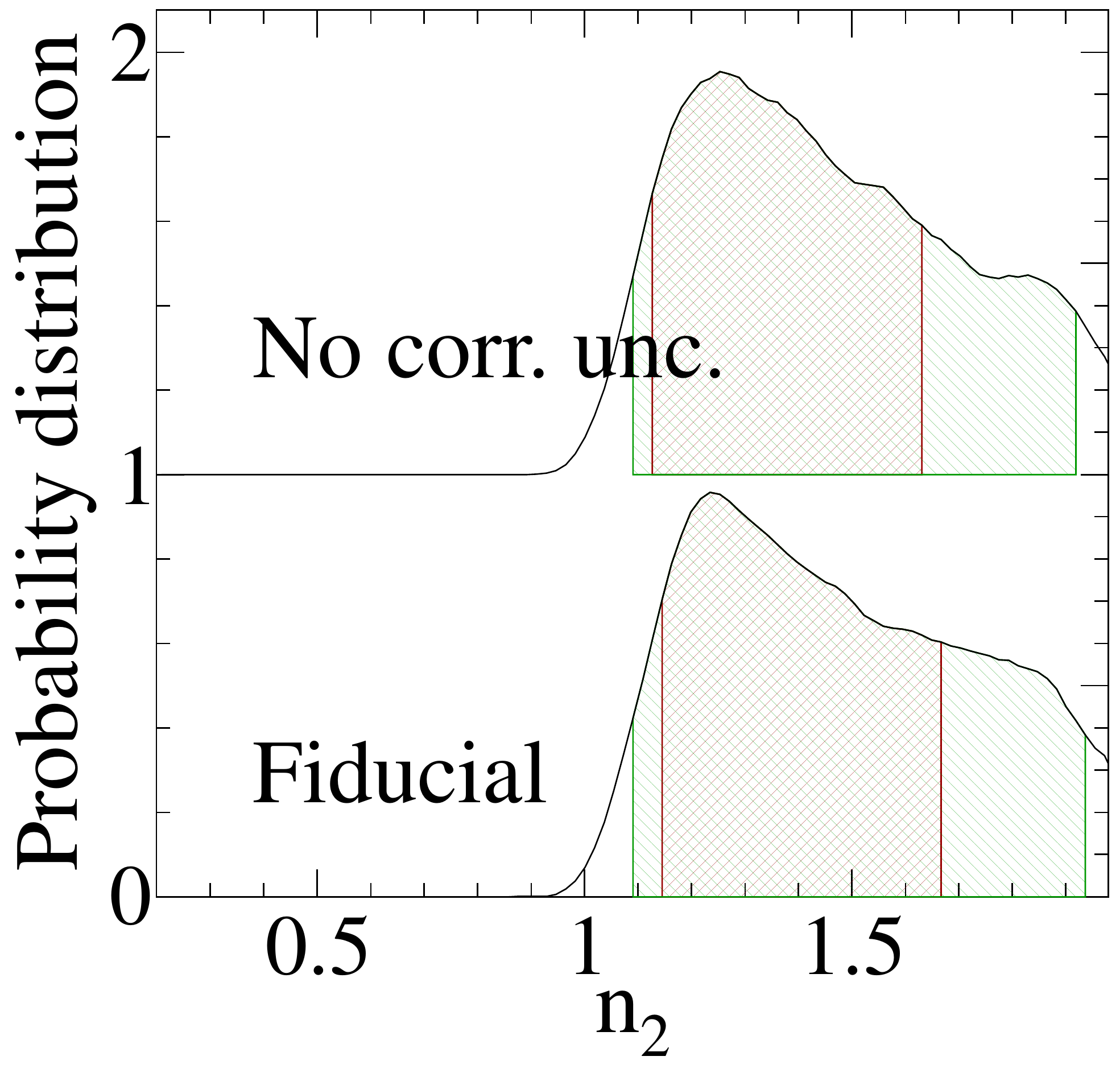}
\caption{The probability distributions and 68\% (dark red areas)
and 95\% (green areas) confidence ranges for parameters of the
high-density EOS when represented by two polytropes. The polytropic
index of the lower-density polytrope is $n_1$ and the polytropic index
of the higher-density polytrope is $n_2$. The transition between these
polytropes takes place at the energy density specified by
$\varepsilon_1$.}
\label{fig:2poly}
\end{figure*}

\begin{figure*}[h]
\includegraphics[width=1.65in]{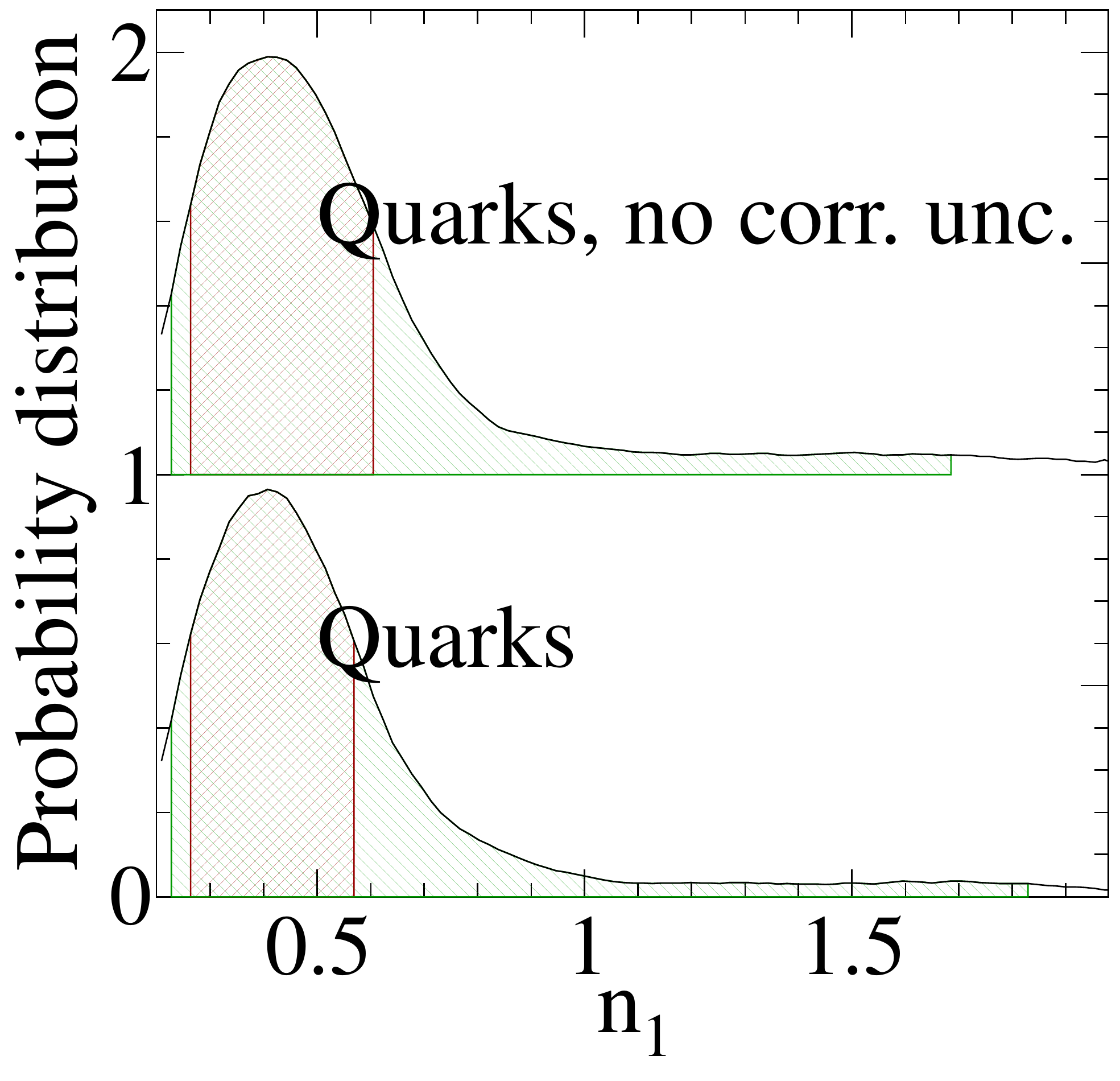}
\includegraphics[width=1.65in]{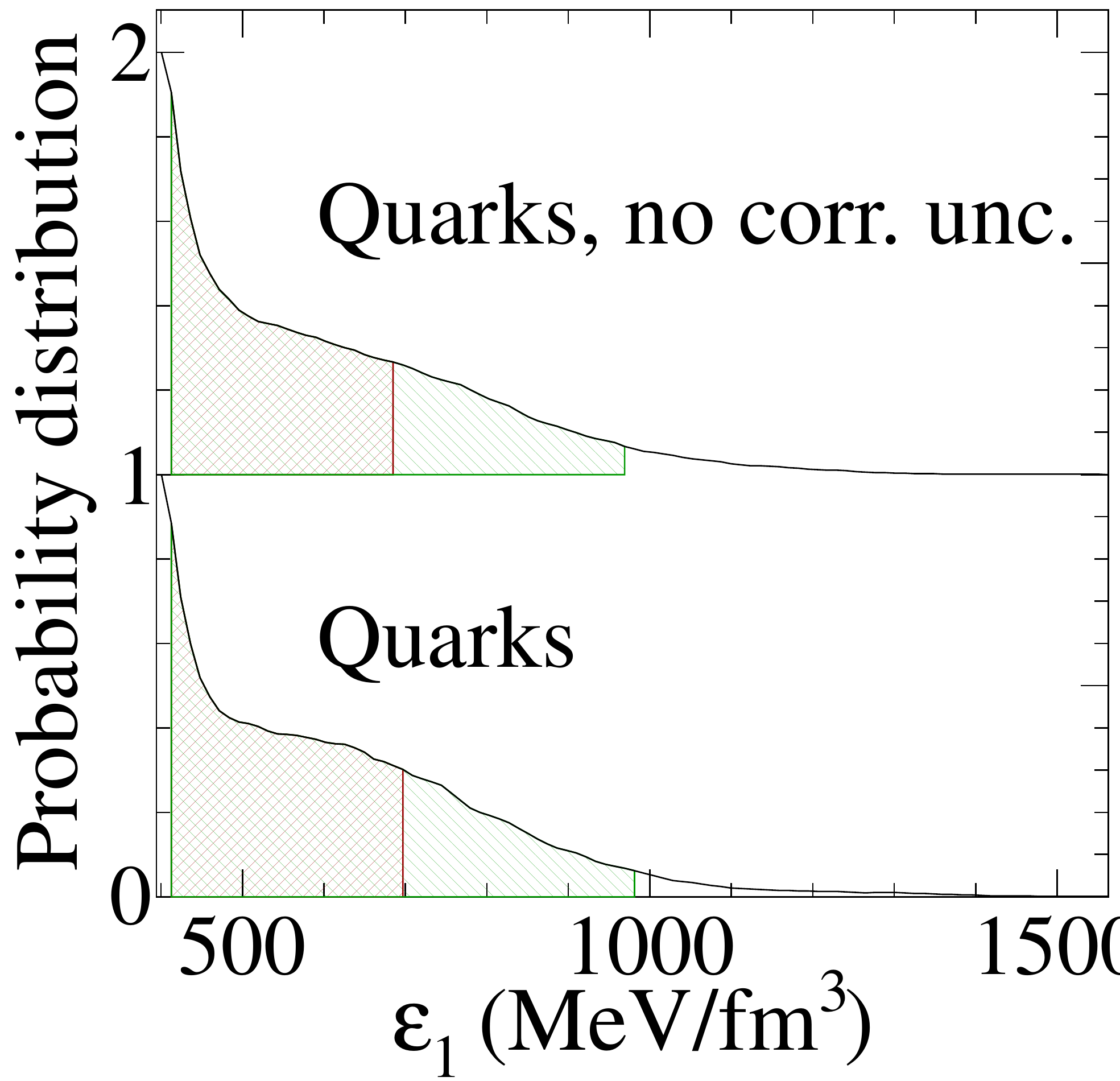}
\includegraphics[width=1.65in]{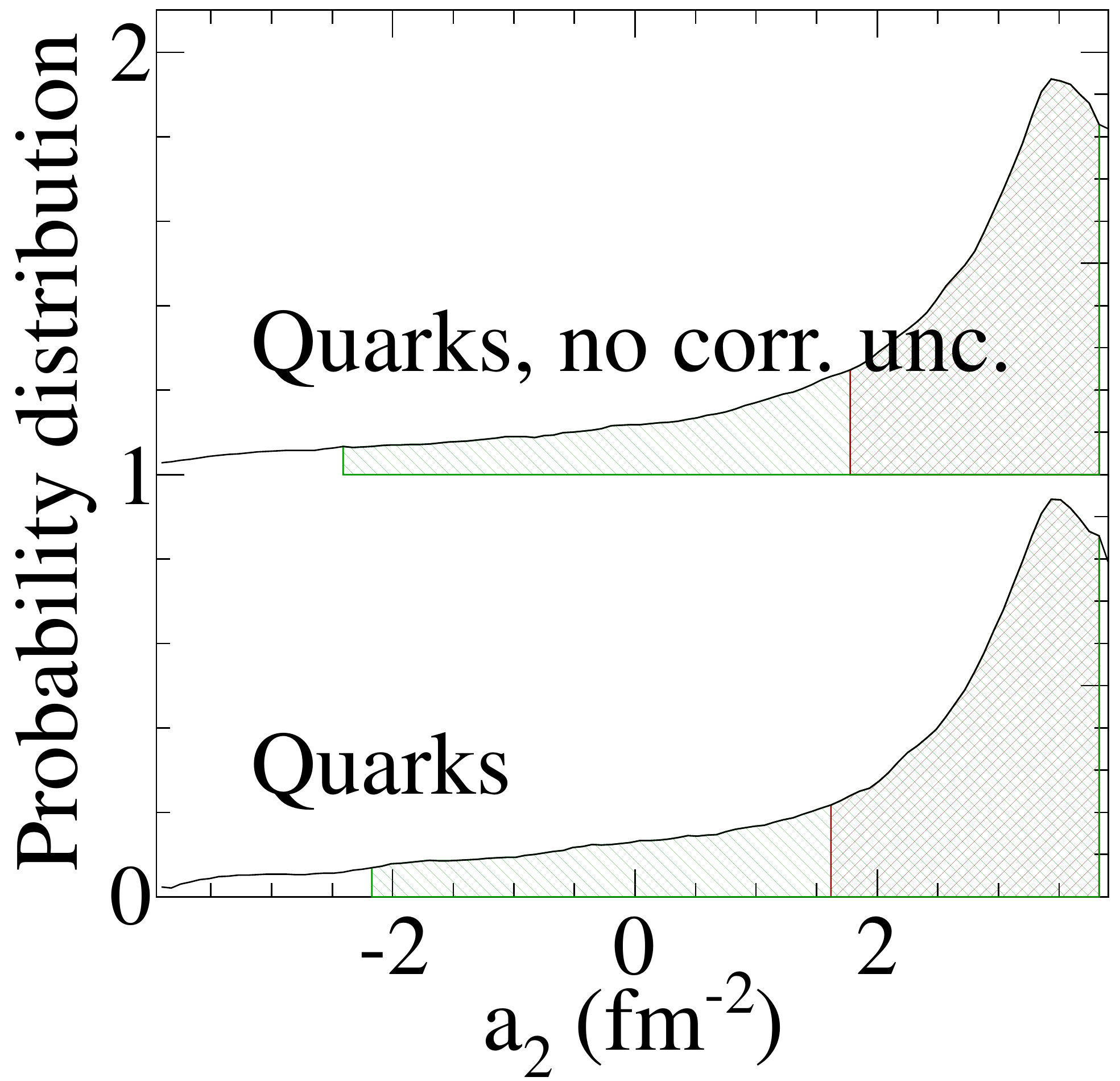}
\includegraphics[width=1.65in]{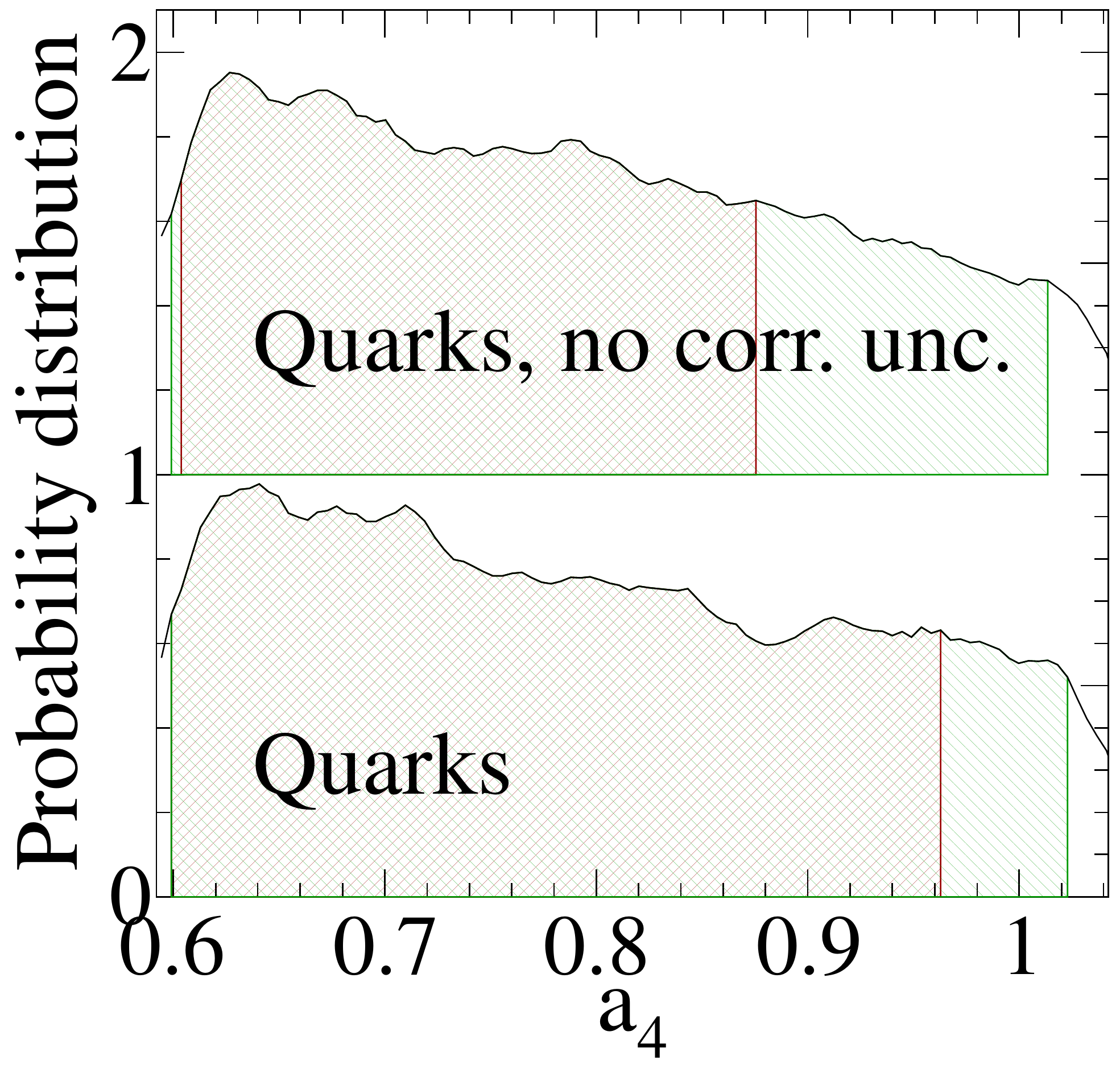}
\caption{The probability distributions and 68\% (dark red areas)
and 95\% (green areas) confidence ranges for parameters of the
high-density EOS when represented by a polytrope (with index $n_1$)
for the mixed phase and deconfined quark matter at higher densities.
The transition between the polytrope and quark matter takes place at
the energy density specified by $\varepsilon_1$. The parameter $a_4$
and $a_2$ represent the coefficients proportional to $\mu^4$ and
$\mu^2$ in the pressure where $\mu$ is the quark chemical potential.}
\label{fig:polyqm}
\end{figure*}

\widetext{
In this supplemental material section we provide the parametrization
of the high-density part of the EOS described in the paper. The two
different parametrizations, using two polytropes or one polytrope and
a quark matter model, are presented in Figs.~\ref{fig:2poly}
and~\ref{fig:polyqm}.}

\vspace{1in}

\vspace{1in}

\end{document}